# *Operando* atomic-scale study of graphene CVD growth at steps of polycrystalline nickel

*Zhiyu Zou[1,2], Virginia Carnevali[3,†], Laerte L. Patera[1,3,‡], Matteo Jugovac[3,§], Cinzia Cepek[1], Maria Peressi[3,\*], Giovanni Comelli[1,3], Cristina Africh[1,\*]*

[1]CNR-IOM, Laboratorio TASC, S.S. 14 km 163.5 - Basovizza, 34149 Trieste, Italy.

[2]Institute of Applied Physics, Vienna University of Technology, 1040 Vienna, Austria.

[3]Department of Physics, University of Trieste, via A. Valerio 2, 34127 Trieste, Italy.

[1] Corresponding Authors. Email: * africh@iom.cnr.it (Cristina Africh, experiments); * peressi@ts.infn.it (Maria Peressi, theory)

† Present Address: Department of Physics, Central Michigan University, Mount Pleasant, Michigan 48858, United States.

‡ Present Address: Institute of Experimental and Applied Physics, University of Regensburg, D-93053 Regensburg, Germany.

§ Present Address: Peter Grünberg Institute (PGI-6), Research Center Jülich, 52425 Jülich, Germany.

**ABSTRACT**

An *operando* investigation of graphene growth on (100) grains of polycrystalline nickel (Ni) surfaces was performed by means of variable-temperature scanning tunneling microscopy complemented by density functional theory simulations. A clear description of the atomistic mechanisms ruling the graphene expansion process at the stepped regions of the substrate is provided, showing that different routes can be followed, depending on the height of the steps to be crossed. When a growing graphene flake reaches a monoatomic step, it extends jointly with the underlying Ni layer; for higher Ni edges, a different process, involving step retraction and graphene landing, becomes active. At step bunches, the latter mechanism leads to a peculiar 'staircase formation' behavior, where terraces of equal width form under the overgrowing graphene, driven by a balance in the energy cost between C-Ni bond formation and stress accumulation in the carbon layer. Our results represent a step towards bridging the material gap in searching new strategies and methods for the optimization of chemical vapor deposition graphene production on polycrystalline metal surfaces.

1. Introduction

Pushed by the expected requirement of large-scale production of high-quality, low-cost films in future electronics and optoelectronics, graphene growth on polycrystalline metals *via* chemical vapor deposition (CVD) has experienced a fast-paced research progress in the past ten years.[1-5] The use of polycrystalline substrates introduces severe complications with respect to single crystals, where the surface has atomically flat morphology and homogeneous crystallinity, since polycrystalline metallic surfaces contain highly stepped regions, step bunches, grain boundaries and even amorphous areas, leading to the well-known 'material gap'. The growth mechanisms on polycrystalline metal surfaces have been investigated at different length scales, from meso- to nano-scale.[6-9] For example, with the aid of mesoscale in-situ monitoring techniques, such as low energy electron microscopy (LEEM) and scanning electron microscopy (SEM), it has been shown that the growth velocity of graphene on polycrystalline transition metals correlates with the relative overlayer-substrate orientation as well as with the grain orientation.[6] LEEM monitoring also reveals that graphene growth can be affected by surface imperfections such as step bunches.[7] Besides, at certain growth conditions, the continuity of graphene can be preserved even across grain boundaries or highly corrugated regions, as observed by post-growth analysis.[8, 10] Despite all these research efforts, no clear atomic scale details on how the growth process develops across these regions have been yet provided.

High-temperature scanning tunneling microscopy (STM) is a suitable approach to the *operando* investigation of surface process,[11-14] which has been adopted to monitor graphene growth on single crystalline transition metal surfaces at various length scales and with different spatial resolution.[12-14] Recently, the feasibility of STM measurements has been proved also for graphene on polycrystalline nickel (Ni-poly), after suitable sample pretreatment.[15] *Operando* STM investigations of graphene growth on polycrystalline substrates appears therefore to be a viable route to clarify the atomistic details of the process.

In this work, we have performed *operando* STM experiments for CVD graphene growth on Ni-poly substrates, one of the most widely used transition metal catalysts for graphene production, focusing in particular on stepped regions. Real-time information of the growth process, providing also atomic-scale details, has been obtained at various stepped regions, including even high step bunches. The experimental results were interpreted with the aid of extensive density functional theory (DFT) simulations, clarifying the role of the interaction between the growing graphene layer and the substrate in the process.[16] A clear description of the graphene growth mechanism on stepped regions is obtained, rationalizing the observed continuity of the layer as due to a very peculiar 'staircase formation' by opening step bunches.

## 2. Methods

### 2.1 Experimental methods

The Ni-poly foils were pretreated according to the procedure previously described.[15] UHV-XPS was adopted to verify that the samples were contaminant-free. Graphene was grown in a UHV chamber (base pressure: $1\times10^{-10}$ mbar) by ethylene exposure ($p$ = $5\times10^{-8}$-$5\times10^{-6}$ mbar) or *via* segregation from residual carbon at 400-500 °C. Imaging was performed with an Omicron variable temperature STM (VT-STM) during growth. An add-on FAST module was used to achieve high frame acquisition speed (10 Hz) in a quasi-constant height mode.[17] Conventional STM scanning was performed in constant current mode. Raw FAST time series were initially processed through a custom-made Python package and an ImageJ plugin. Each single STM frame was processed by Gwyddion.[18]

**2.2 Theoretical methods**

DFT calculations were performed with the Quantum ESPRESSO code,[19] using plane-wave basis set and the Generalized Gradient Approximation for the exchange-correlation functional in the Perdew-Burke-Ernzerhof parametrization (GGA-PBE).[20] After convergence test, the kinetic energy cutoff for the plane-wave basis set was set at 30 Ry. The equilibrium lattice parameters characterizing the clean Ni(100) surface and the free-standing graphene are $a_{Ni}= \overline{a_{bulk}}$ = 2.49 Å and $a_{GR}$ = 2.46 Å respectively, equal to the experimental values. In order to describe the graphene/Ni(100) interaction, the van der Waals correction term was included using the semiempirical DFT-D approach.[21] As largely adopted for DFT surface studies, slab models have been used for Ni surfaces, with 3 Ni layers and graphene overlayer on one side only, and a vacuum spacing of about 15 Å between graphene and the consecutive surface of the repeated parallel slab. We focused on monoatomic stepped surfaces, which are particular vicinal surfaces (i.e., surfaces cut at an angle slightly different to that of the (100) surface). Explicitly, the (11N) vicinal surface with even N corresponds to a stepped monoatomic surface with (100) terraces of constant width equal to N/2 $a_{Ni}$. Our simulation cells are orthorhombic and always contain two terraces in order to account for the periodicity of the graphene overlayer along its armchair direction (see Fig. S8 in SI). The dimension of the simulation cells along the step edges ([011] direction) is therefore $a_{Ni}$, whereas it depends on N in the other in-plane direction, [  ]. Due to the large cell dimensions, a 1×2×1 Monkhorst-Pack k-point grid centered on the Gamma point has been used for Brillouin zone integration,[22] with an energy broadening of 0.01 Ry.[23] Only the nickel layer at the surface opposite to graphene was kept fixed, and all the other atomic positions were optimized by total energy and forces minimization. The simulations of the graphene ribbons with different width have been performed using orthorhombic cells with a (100) terminated Ni slab, with length equal to 11 $a_{Ni}$ in the direction of the growing ribbon. Stick-and-ball models were rendered with the VMD software.[24] STM images were obtained using the Tersoff-Hamann approach[25] simulating the constant current experimental detection mode. Whereas in the experimental setup the STM tip is perpendicular to the terraces, in the simulation cell the terraces are slightly tilted with respect to the cell basis vector which mimics the acquisition direction: this originated a fictitious gradient in the simulated STM images which was therefore *a posteriori* corrected with Gwyddion software.[18]

## 3. Results and discussion

### 3.1 Joint growth mechanism of graphene together with monoatomic step edge of nickel

The same series of simple pre-treatments already described in our previous work[15] was used to prepare Ni-poly foils suitable for STM measurements. Here, we focus on grains exposing the (100) facet, one of the ubiquitous orientations in Ni-poly substrates.[15, 26-27] The extended moiré pattern which originates on these facets from the interface symmetry mismatch can serve as a hallmark of graphene formation and a magnification of the defects in graphene lattice even at a relatively large scan scale.[15, 28-30] It should be noted that this experimental approach could be easily extended to differently oriented facets in Ni and even other polycrystalline transition metal foils, whenever graphene grows at a temperature and a hydrocarbon pressure compatible with the specifics of the microscope.

The (100) facet of the Ni-poly substrate is found to be featured with flat terraces as wide as tens of nanometers, separated by step bunches which can span up to tens of atomic layers.[15] It should be noted that with the growth conditions adopted in this work (see Methods), graphene is always monolayer, as recently found also on Ni(100) thin films *via* ambient-pressure CVD preparation;[29] moreover, like the case of Ni(111) single crystals, graphene never grows rightly above nickel carbide, at variance with the behavior observed on early transition metals.[31-32] Similarly to what already reported for the Ni(111) single crystal,[11] two graphene growth mechanisms are active on flat (100) terraces of Ni-poly, corresponding to in-plane transformation and on-terrace adlayer growth, respectively, which will be reported elsewhere. In the following we will focus on the peculiar growth behavior observed at steps.

Notably, growth dynamics at step edges is expected to play an important role in the overall graphene CVD process under industrial conditions, since polycrystalline metal surfaces are characterized by the presence of step bunches.[7, 33]

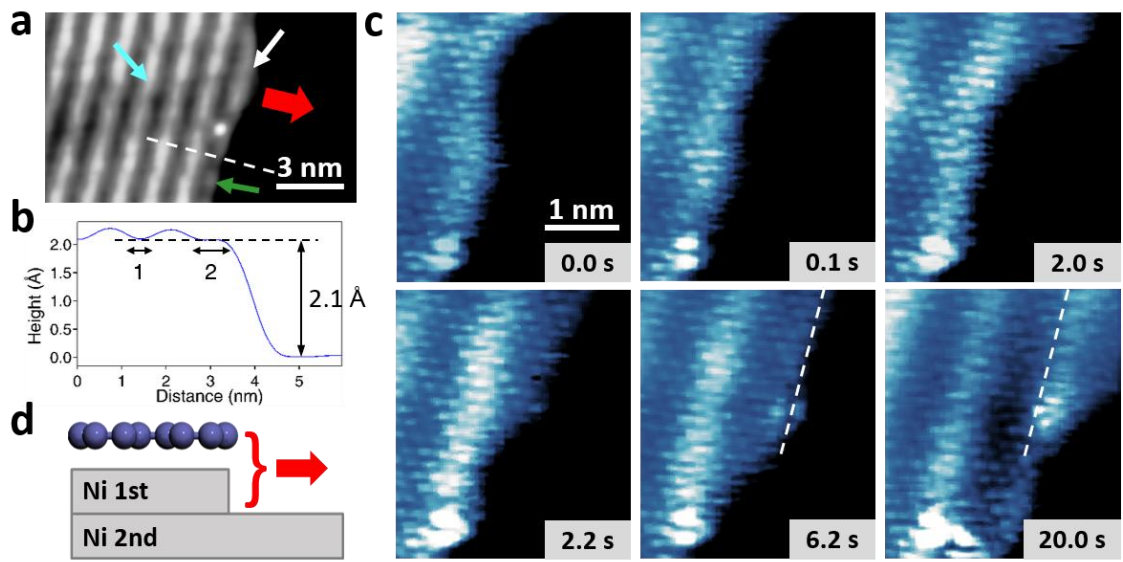

**Figure 1.** Graphene growth mechanism at a polycrystalline Ni monoatomic step edge. (a) Constant-current STM image of a growing graphene flake at a single step edge. (b) Line profile along the white dashed line in (a). Segments 1 and 2 mark the estimated chemisorbed regions of graphene on flat terrace and step edge, respectively. (c) Selected frames from a Fast-STM movie (frame rate: 10 Hz, $V = -20$ mV, $I = 7$ nA, see corresponding Movie S1) acquired at the edge location in (a) in *quasi*-constant height mode. The misorientation angle of graphene from the Ni(100) lattice is $\theta \sim 6.5°$. The white dashed lines along the moiré stripes for the 6.2 s and 20.0 s frames are guides to the eye, and correspond to the same positions. (d) Schematic model qualitatively illustrating the joint growth mechanism. The red arrow shows the joint growth direction of graphene and Ni 1st layer.

Fig. 1a shows the STM image of a graphene flake at a monoatomic Ni step, as evidenced by the apparent height difference between the growing layer on the left and the Ni terrace on the right (panel b), which is larger than the apparent height of a bare monoatomic step (~1.7 Å). The graphene flake, which completely covers the upper terrace, shows the typical striped moiré previously observed on flat (100) facets. As discussed in Ref. 13, the stripes with alternating physi- and chemisorbed regions originate from the symmetry and lattice mismatch between graphene and Ni(100) for small misorientation angle $\theta$, defined as the smallest angle between one zigzag direction of graphene and the Ni(100) lattice vectors. Here the growth process was captured by Fast-STM image sequences in *quasi*-constant height scanning mode, with a frame rate up to 10 Hz;[34] selected frames of the resulting movie (Movie S1) are shown in panel c. We notice that as the growth proceeds, the edge of the graphene/nickel growing front remains always sharp, in strong contrast to the fuzzy termination observed for growing graphene adlayers on flat terraces (to be reported elsewhere). This suggests that, when diffusing nickel adatoms are trapped at the metal edge, they are immediately covered and stabilized by the simultaneously growing graphene.

This mutual stabilization of graphene and Ni at the edge is also corroborated by the following analysis of the moiré corrugation close to the edge. While in the middle of the terrace (cyan arrow in panel a) bright stripes/dark troughs of moiré patterns display an almost constant width along their direction, larger flat graphene regions are imaged at the step (green arrow, see also the line profile in panel b). The regular buckling period is restored only upon further extension of the growing region (white arrow). As will be discussed later, the observed behavior is related to the balance between the strength of the graphene-Ni bonding and the accumulated stress in the layer. Such observation is confirmed by the Fast-STM frames shown in Fig. 1c, in particular by comparing the corrugations next to the white dashed lines in panels 6.2 s and 20.0 s.

Images in Fig. 1c also reveal that the growth process proceeds in an irregular way both in time and space (see Fig. S1 and Movie S1 for details). More specifically, the growth velocity of graphene along the direction of moiré stripes (white dashed line) is much faster than in the direction normal to it. Furthermore, the carbon addition rate significantly varies during the process, pointing to a relevant role of the specific atomic configuration, determining the active sites for the joint growth (see SI for more details).

Summarizing, the growth processes observed so far proceed as follows: when the graphene growing front reaches a monoatomic nickel step, the strong interface bonding determines a joint growth of both layers, rather than an uncorrelated growth, which otherwise would lead to either downhill flow of graphene over the edge or forward movement of the bare Ni step.

**3.2 Downhill landing of graphene over monoatomic step edge of nickel**

The situation significantly changes when the graphene edge reaches a multiple nickel step. As shown in Fig. 2, in this case it is required, for the growth to proceed, that the step height is reduced. Therefore, the graphene flake extends in a peculiar way: the multiple step opens by the 'landing after step retraction' process shown in Fig. 2 (see also Movies S2,3 and Fig. S2). The topmost Ni layer under graphene in Fig. 2a loses Ni atoms that diffuse away from the edge (step retraction), forcing the flake to fall down by one floor (landing), similarly to what previously reported for the growth of graphene on monoatomic steps of ruthenium.[13] The graphene layer then grows together with the remaining Ni bilayer (panels b and c), in a process similar to the monoatomic case described above. However, in this case the joint advance of graphene with the two Ni layers is a process more difficult than the retraction of the new topmost Ni layer (panel d), which leads to graphene finally lying on a monoatomic step, and the growth can then easily follow the joint growth mechanism reported in Fig. 1 (panel e).

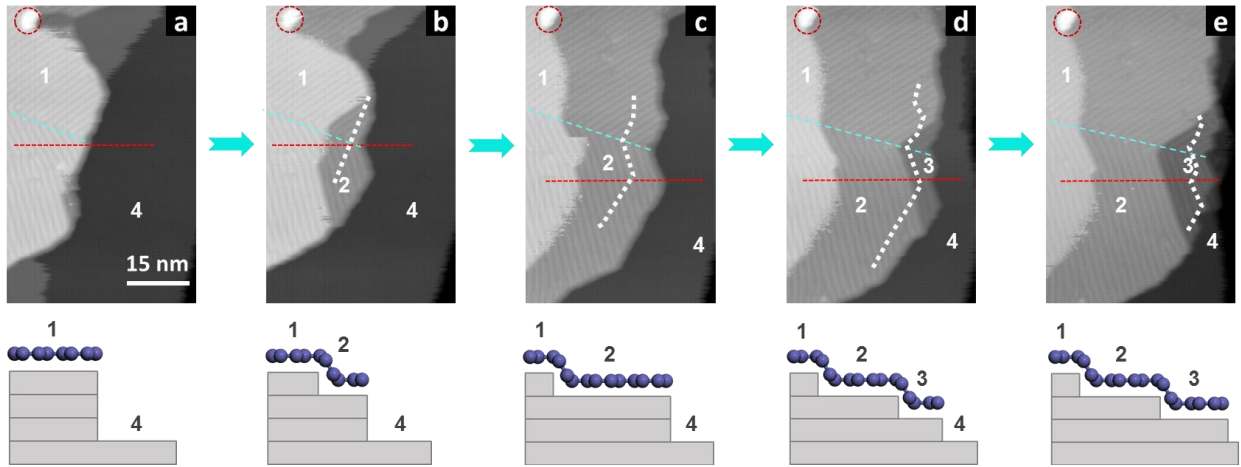

**Figure 2.** Constant-current STM images (upper panels) and corresponding qualitatively illustrated models (lower panel, side-views along the red dashed lines in the upper panel) of graphene growth at a multiple-stepped area. Images were acquired during growth at: 0 s (a), 42 s (b), 125 s (c), 170 s (d) and 212 s (e). In the upper panels, the cyan dashed lines mark the graphene domain boundary, while the white dotted lines in stages b-e mark the growth frontiers in their previous stages a-d. The feature marked by the red circle is used to calibrate the position in STM images.

Noteworthy, the newly formed borders between upper and lower graphene (such as borders between 1/2 in panel b and 2/3 in panel d) are parallel to the moiré stripes, indicating that the periodically modulated interface interaction affects the downhill landing process. Likely, moving nickel atoms under the physisorbed region of the moiré structure costs less energy than under the chemisorbed area, which may result in anisotropic Ni diffusion and in step edges aligned with the moiré stripes (see also Figs. 1a,c).

### 3.3 Graphene overgrowth at step bunches of polycrystalline nickel

Increasing the complexity of the step structure, after the characterization of graphene growth at monoatomic and multiple steps, we have studied the case of high step bunches. This is particularly relevant, as step bunches widely exist on polycrystalline metal foils from the manufacturing procedures.[7, 33] Besides, the presence of highly stepped regions could influence the continuity of graphene,[16] and thus its electronic performance.

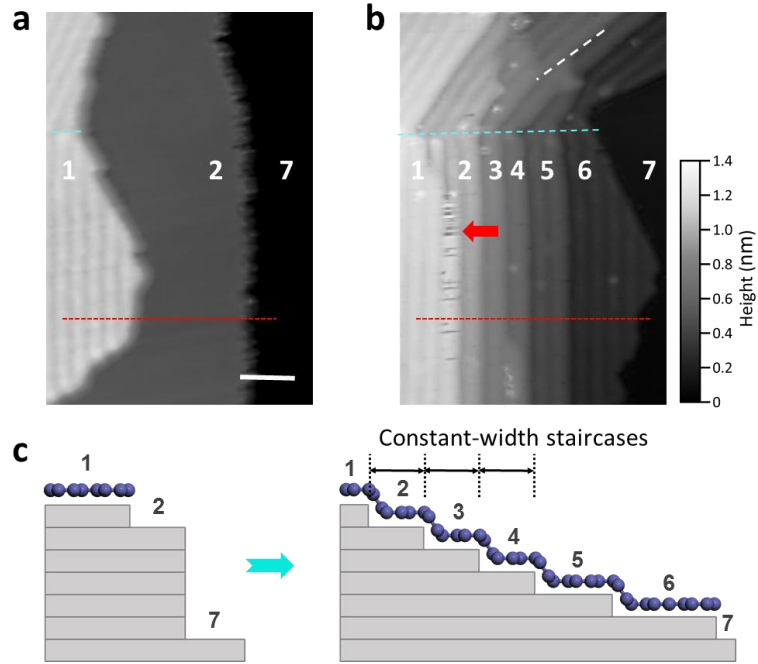

**Figure 3.** Staircase formation at a step bunch, leading to a series of consecutive constant-width terraces separated by monoatomic step edges. (a,b) Constant-current STM images of graphene growth at a step bunch, acquired at 0 s (a) and 170 s (b). (c) Schematic models qualitatively illustrating the process (side view) along the red dashed lines in (a) and (b). The vertical direction in panels a and b corresponds to ~0° misorientation of the moiré.

The process at step bunches, visualized in Fig. 3, is based on the one active for multilayer steps. The first frame (Fig. 3a) shows on the left a graphene flake supported by one Ni(100) atomic layer, growing towards the fuzzy step bunch on the right separating two nickel terraces, labeled 2 (upper) and 7 (lower) (see left illustration in Fig. 3c). The line profile in Fig. S3 indicates that the step bunch corresponds to 5 Ni(100) atomic layers. When the graphene flake reaches the step bunch, the latter is overcome by opening the bunch in a sequence of monoatomic steps *via* the 'landing after step retraction' mechanism described above. After 170 s (Fig. 3 (b) and S4), the step bunch has separated, forming 4 narrow (areas 2-5) and 1 wider (area 6) terraces, all covered by graphene, while the lowest Ni terrace (7) is still uncovered. The linear graphene domain boundary (cyan dashed line) between the moiré superstructures with two different orientations indicates that the film is continuous. Interestingly, at variance with the case of Cu(100), where the growth frontiers of graphene can be severely distorted by the step bunches,[7] here the moiré stripes orientation and the step edges of the Ni(100) substrate are always aligned.

A remarkable feature of the process described above is that the consecutive narrow terraces generated by this process typically have a constant width (terraces 2-4), corresponding to about 1.4-1.5 times the width of the moiré beatings on flat terraces, as shown in Fig. 4a,b. The staircase formation mechanism appears to be the same regardless of the specific misalignment angle of graphene (see Figs. S4-6), although the absolute values of the period of the moiré beatings on flat terraces and of the staircase terrace width vary for different angles.

**3.4 DFT simulation of constant-width terraces at step bunches**

In order to rationalize the observed behavior, a series of DFT calculations for the graphene aligned with the substrate (0° misorientation angle) has been performed. First of all we tried to clarify the origin of the previously described moiré pattern adopted by an infinite graphene layer on Ni(100).[15] To this purpose, we simulate a growing graphene flake with ribbons of different width in the growing [   ] direction and infinitely long along the moiré stripes parallel to the [011] direction (Fig. 4c, see Fig. S7 for more details). One zigzag edge is fixed at the position corresponding to a moiré valley, mimicking the chemisorbed region closest to the growing front, whereas the rest of the ribbon is left free to find its preferential structure. The addition of new carbon atoms to the graphene flake is simulated by increasing the ribbon width along the armchair direction. According to our calculations, the ribbons remain flat and chemisorbed to the substrate up to a critical width of about thirteen carbon rows; larger ribbons start corrugating, assuming a striped moiré motif (Figs. 4c, S7), which characterizes also the infinite layer (Fig. 4d). This behavior can be explained by the co-existence of two competing mechanisms, involving bonding and stress: on the one side, the graphene network is stretched to keep an optimal registry with the substrate, maximizing the bonding with the substrate; on the other hand, this process induces in graphene a progressive negative stress that needs to be sustained. When the latter exceeds a certain threshold, the graphene network recovers an unstrained (unstressed) configuration, and forms regular wiggles separating the chemisorbed regions, which can be corroborated by the observation in Fig. 1a. Based on our simulations, the maximum excess stress in the optimized relaxed graphene ribbons is about -10 kbar, which thus is an estimate of the maximum stress in a stretching situation. From our investigation of infinite graphene layers in s-moiré pattern (see SI), the maximum stress in a compressive situation is about +10 kbar. In case of infinite graphene layers, configurations with even higher positive stress could be forced to exist by imposing periodic boundary conditions, but they are not compatible with experimental images. In conclusion, 10 kbar is a reasonable estimate of the maximum stress value, either tensile or compressive, in reliable models of this system.

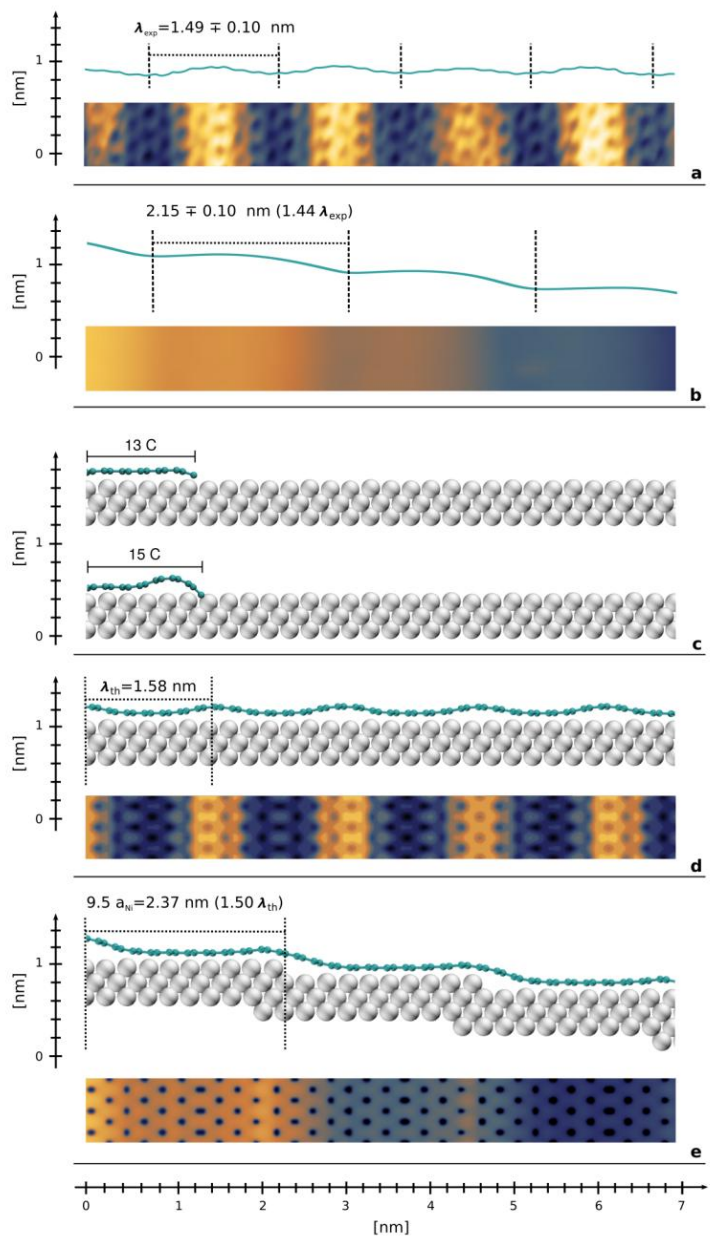

**Figure 4.** (a,b) Experimental constant-current STM images of the graphene moiré on a flat (a) and a stepped (b) Ni(100) regions covered by a graphene flake. (c) DFT optimized models of two graphene ribbons with different width on a flat Ni(100) surface, where one end (left) was fixed. (d) DFT model of an infinite striped moiré on a flat Ni (100) surface, together with its corresponding simulated STM image. (e) DFT model of a graphene overlayer carpeting a stepped Ni(100) surface composed of monoatomic

terraces of 9.5 $a_{Ni}$ width with minimum stress and maximum bonding to the substrate, and its simulated STM image.

The simulation of graphene growth on stepped surfaces is a much more complicated task, thus we only compare different configurations of terraces covered by graphene overlayers. We focus on sequences of terraces of equal width separated by monoatomic steps, which already require quite large simulation cells (Fig. S8). Four different cell configurations have been simulated, characterized by two Ni terraces with a total width of 15, 17, 19 and 21 $a_{Ni}$ (Ni-Ni bond length, 2.49 Å) respectively, and a number of C rows maintaining an almost constant ratio of ~2.35±0.05 with the number of surface Ni rows (see Fig. S9b-e and Methods for details). In all these cases, a smooth regular carpeting of graphene is found, lying flat over the terraces, with a periodicity longer than the moiré period on the flat surface.[15] The average C-C distance, which varies by less than 3% among the different cases, quantifies the strain, whereas the average C-Ni distance, which varies by about 3%, could be tentatively considered as an estimate of the graphene-substrate bonding strength (the shorter the distance, the stronger the bonding). The excess stress due to the graphene overlayer, stretched or compressed with respect to an ideal unstressed situation, varies from -11 kbar to +18 kbar. Two of these configurations (b and c in Fig. S9) are characterized by an excess stress above the threshold of 10 kbar reported above as the maximum stress that can be compensated by the formation of covalent bonds with the substrate, and thus should be considered as artificial solutions due to the imposed boundary conditions. Between the remaining two, the configuration in Fig. S9d shows the smallest stress and the shortest average C-Ni distance (indicative of a stronger C-Ni bonding), and thus is expected to be the best candidate model for the observed structure. This configuration, characterized by terraces of 9.5 $a_{Ni}$ width, corresponding to 1.5 times the beating of the striped moiré on the flat (100) surface, is also shown in Fig. 4e together with its simulated STM image. The reported absolute values of the periodicities of the graphene superstructures are affected by specific parameters and factors both for the experiment (e.g. sample temperature, thermal drifts and image calibration) and simulations (e.g. cell size), making a direct comparison not very meaningful. However, when comparing the experimental and theoretical ratios between the staircase terrace width and the period of the moiré beatings on flat terraces most of these effects cancel out. Indeed, the 1.5 ratio obtained by numerical simulations well matches the experimental value (1.44), supporting the conclusion that the driving force behind the observed staircase formation of high steps under the growing graphene sheet is the formation of monoatomic steps between terraces with the ideal width to optimize the energy cost balance between stress and bond formation. The observed behavior is opposite of the one suggested for Cu(111), where graphene growth is reported to induce step bunching.[35] This can be rationalized in terms of the much stronger chemical bonding established by C atoms with the Ni substrate in our case and questions the existence of a general step bunching mechanism working for different substrates. It appears therefore that significant differences exist between the growth mechanism of graphene on Cu and Ni, the latter being considered as a promising alternative candidate for industrial production of large graphene foils.

Finally, we note that recently the role of vicinal steps in the nucleation of 2D monocrystals of hexagonal boron nitride (h-BN) was discussed [NL, Nature], highlighting their effect in determining the orientation of the resulting flake. At variance with what we observe for graphene, which is found to grow across even high step edges without the need for new nucleation, thus yielding a continuous flake across the surface, for h-BN the high steps are found to constitute a high-energy barrier, making the spillover over the step prohibitive. The growth mechanism therefore appears to be very different in the two cases.

## 4. Conclusion

To summarize, we characterized the atomistic growth mechanisms of graphene on technologically-relevant highly-stepped polycrystalline Ni surface, by means of *operando* STM studies up to video rate. On corrugated regions, the growth process requires a substantial rearrangement of the substrate, involving peculiar step dynamics. At monoatomic Ni steps, the strong metal-carbon interface bonding gives rise to a simultaneous extension of the graphene layer and the underlying Ni terrace. At higher steps, this process competes with step retraction, allowing graphene to land on the lower Ni terrace. Finally, at step bunches, the need to release the stress of graphene while maximizing the C-Ni bonding imposes to open a staircase with the formation of substrate terraces of specific widths, as nicely corroborated by DFT calculations. Graphene growth leads thus to a general smoothing of the underlying metal substrate, exploiting the high mobility of Ni atoms under the graphene cover.

Our results rationalize the continuity of CVD-graphene on highly stepped substrates and show the possibility of growing graphene single crystals on rough polycrystalline metal surfaces with a strong interface interaction without being limited by the presence of high steps in the substrate.

Our work provides new insight towards bridging the 'material gap' between application-relevant CVD production and the UHV surface-science exploration of graphene, and demonstrates the possibility of using polycrystalline surfaces with relatively high corrugation as a new playground for *operando* scanning probe microscopy measurements.

**Appendix: Supporting Information**

Detailed STM image and movie analysis, details of DFT simulations (pdf file)

STM movies Movie_S1, Movie_S2, Movie_S3, Movie_S4 (mp4 files)

**Author Contributions**

The manuscript was written through contributions of all authors. All authors have given approval to the final version of the manuscript.

**ACKNOWLEDGMENTS**

Z.Z. acknowledges support by the 'ICTP TRIL Programme, Trieste, Italy' in the framework of the agreements with the Elettra and CNR-IOM laboratories. We acknowledge support from: EU-H2020 Research and Innovation programme under grant agreement no. 654360 Nanoscience Foundries and Fine Analysis - Europe; Italian Ministry of Foreign Affairs and International Cooperation (MAECI) for the Project PGR06529; University of Trieste for FRA2018. Computational resources have been obtained from the CINECA Italian Consortium of Universities through the ISCRA initiative and the agreement with the University of Trieste.

**REFERENCES**

[1]   Q.K. Yu, J. Lian, S. Siriponglert, H. Li, Y.P. Chen, S.-S. Pei, Graphene segregated on Ni surfaces and transferred to insulators, Appl. Phys. Lett. 93 (2008) 113103

[2]   A. Reina, X.T. Jia, J. Ho, D. Nezich, H. Son, V. Bulovic et al, Large area, few-layer graphene films on arbitrary substrates by chemical vapor deposition, Nano Lett. 9 (2009) 30-35

[3]   X.S. Li, W.W. Cai, J.H. An, S. Kim, J. Nah, D.X. Yang et al, Large-area synthesis of high-quality and uniform graphene films on copper foils, Science 324 (2009) 1312-1314

[4]   B.Y. Dai, L. Fu, Z.Y. Zou, M. Wang, H.T. Xu, S. Wang et al, Rational design of a binary metal alloy for chemical vapour deposition growth of uniform single-layer graphene, Nat. Commun. 2 (2011) 522-527

[5]   Z.Y. Zou, B.Y. Dai, Z.F. Liu, CVD process engineering for designed growth of graphene, Sci. China Chem. 43 (2013) 1-17

[6]   R.S. Weatherup, A.J. Shahani, Z.J. Wang, K. Mingard, A.J. Pollard, M.-G. Willinger et al, In situ graphene growth dynamics on polycrystalline catalyst foils, Nano Lett. 16 (2016) 6196-6206

[7]  J.M. Wofford, S. Nie, K.F. McCarty, N.C. Bartelt, O.D. Dubon, Graphene Islands on Cu Foils: The Interplay between Shape, Orientation, and Defects, Nano Lett. 10 (2010) 4890-4896

[8]  H.I. Rasool, E.B. Song, M.J. Allen, J.K. Wassei, R.B. Kaner, K.L. Wang et al, Continuity of graphene on polycrystalline copper, Nano Lett. 11 (2011) 251-256

[9]  H. Kim, C. Mattevi, M.R. Calvo, J.C. Oberg, L. Artiglia, S. Agnoli et al, Activation energy paths for graphene nucleation and growth on Cu, ACS Nano 6 (2012) 3614-3623

[10] R. Zhao, Y. Zhang, T. Gao, Y. Gao, N. Liu, L. Fu et al, Scanning tunneling microscope observations of non-AB stacking of graphene on Ni films, Nano Res. 4 (2011) 712-721

[11] L.L. Patera, C. Africh, R.S. Weatherup, R. Blume, S. Bhardwaj, C. Castellarin-Cudia et al, In situ observations of the atomistic mechanisms of Ni catalyzed low temperature graphene growth, ACS Nano 7 (2013) 7901-7912

[12] L.L. Patera, F. Bianchini, C. Africh, C. Dri, G. Soldano, M.M. Mariscal et al, Real-time imaging of adatom-promoted graphene growth on nickel, Science 359 (2018) 1243-1246

[13] S. Günther, S. Dänhardt, B. Wang, M.L. Bocquet, S. Schmitt, J. Wintterlin, Single terrace growth of graphene on a metal surface, Nano Lett. 11 (2011) 1895-1900

[14] S. Günther, S. Dänhardt, M. Ehrensperger, P. Zeller, S. Schmitt, J. Wintterlin, High-Temperature Scanning Tunneling Microscopy Study of the Ordering Transition of an Amorphous Carbon Layer into Graphene on Ruthenium(0001), ACS Nano 7 (2013) 154-164

[15] Z.Y. Zou, V. Carnevali, M. Jugovac, L.L. Patera, A. Sala, M. Panighel et al, Graphene on nickel (100) micrograins: Modulating the interface interaction by extended moiré superstructures, Carbon 130 (2018) 441-447

[16] A. Ouerghi, M.G. Silly, M. Marangolo, C. Mathieu, M. Eddrief, M. Picher et al, Large-Area and High-Quality Epitaxial Graphene on Off-Axis SiC Wafers, ACS Nano 6 (2012) 6075-6082

[17] F. Esch, C. Dri, A. Spessot, C. Africh, G. Cautero, D. Giuressi et al, The FAST module: An add-on unit for driving commercial scanning probe microscopes at video rate and beyond, Rev. Sci. Instrum. 82 (2011) 053702

[18] D. Necas, P. Klapetek, Gwyddion: an open-source software for SPM data analysis, Cent. Eur. J. Phys. 10 (2012) 181–188

[19] P. Giannozzi, S. Baroni, N. Bonini, M. Calandra, R. Car, C. Cavazzoni et al, QUANTUM ESPRESSO: a modular and open-source software project for quantum simulations of materials, J. Phys.: Condens. Matter 21 (2009) 395502

[20] J.P. Perdew, K. Burke, M. Ernzerhof, Generalized gradient approximation made simple, Phys. Rev. Lett. 77 (1996) 3865-3868

[21] S. Grimme, Density functional theory with London dispersion corrections, Wiley Interdiscip. Rev.: Comput. Mol. Sci. 1 (2011) 211-228

[22] H.J. Monkhorst, J.D. Pack, Special points for Brillouin-zone integrations, Phys. Rev. B 13 (1976) 5188-5192

[23] M. Methfessel, A.T. Paxton, High-precision sampling for Brillouin-zone integration in metals, Phys. Rev. B 40 (1989) 3616-3621

[24] W. Humphrey, A. Dalke, K. Schulten, VMD: Visual molecular dynamics, J. Mol. Graphics 14 (1996) 33-38

[25] J. Tersoff, D.R. Hamann, Theory of the scanning tunneling microscope, Phys. Rev. B 31 (1985) 805-813

[26] R.S. Weatherup, B.C. Bayer, R. Blume, C. Ducati, C. Baehtz, R. Schlögl et al, In situ characterization of alloy catalysts for low-temperature graphene growth, Nano Lett. 11 (2011) 4154-4160

[27] L. Hyonik, L. Seulah, H. Juree, L. Sang Geun, L. Jae-Hong, L. Taeyoon, Graphene converted from the photoresist material on polycrystalline nickel substrate, Jpn. J. Appl. Phys. 51 (2012) 06FD17

[28] D. Usachov, A.M. Dobrotvorskii, A. Varykhalov, O. Rader, W. Gudat, A.M. Shikin et al, Experimental and theoretical study of the morphology of commensurate and incommensurate graphene layers on Ni single-crystal surfaces, Phys. Rev. B 78 (2008) 085403

[29] D.L. Mafra, J.A. Olmos-Asar, F.R. Negreiros, A. Reina, K.K. Kim, M.S. Dresselhaus et al, Ambient-pressure CVD of graphene on low-index Ni surfaces using methane: A combined experimental and first-principles study, Phys. Rev. Mater. 2 (2018) 073404

[30] J. Coraux, A.T. N'Diaye, C. Busse, T. Michely, Structural coherency of graphene on Ir(111), Nano Lett. 8 (2008) 565-570

[31] Z.Y. Zou, L. Fu, X.J. Song, Y.F. Zhang, Z.F. Liu, Carbide-forming groups IVB-VIB metals: A new territory in the periodic table for CVD growth of graphene, Nano Lett. 14 (2014) 3832-3839

[32] Z.Y. Zou, X.J. Song, K. Chen, Q.Q. Ji, Y.F. Zhang, Z.F. Liu, Uniform single-layer graphene growth on recyclable tungsten foils, Nano Res. 8 (2015) 592-599

[33] S.J. Chae, F. Güneş, K.K. Kim, E.S. Kim, G.H. Han, S.M. Kim et al, Synthesis of Large-Area Graphene Layers on Poly-Nickel Substrate by Chemical Vapor Deposition: Wrinkle Formation, Adv. Mater. 21 (2009) 2328-2333

[34] C. Dri, M. Panighel, D. Tiemann, L.L. Patera, G. Troiano, Y. Fukamori et al, The new FAST module: a portable and transparent add-on module for time-resolved investigations with commercial scanning probe microscopes, Ultramicroscopy 205 (2019) 49-56

[35] D. Yi, D. Luo, Z. Wang, Jun, J. Dong, Chen, X. Zhang, M.-G. Willinger et al, What Drives Metal-Surface Step Bunching in Graphene Chemical Vapor Deposition?, Phys. Rev. Lett. 120 (2018) 246101